\documentclass[a4paper,12pt]{article}

\title{Sources of FCNC in $SU(3)_C \otimes SU(3)_L \otimes U(1)_X$ models}
\author{J. M. Cabarcas\footnote{josecabarcas@usantotomas.edu.co},\\
\it \small{Departamento de Ciencias B\'asicas}\\
\it \small{Universidad Santo Tomas}\\
J. Duarte\footnote{jaduartec@unal.edu.co} and J-Alexis Rodriguez\footnote{jarodriguezl@unal.edu.co}\\
{\it \small{Departamento de F\'isica}} \\ {\it \small{Universidad Nacional de Colombia, Bogot\'a, Colombia}}}

\date{}
\begin{document}
\maketitle

\begin{abstract}
There are different models based on the gauge symmetry $SU(3)_C \otimes SU(3)_L \otimes U(1)_X$ (331)
some of them includes exotic particles and others are constructed without any exotic charges assigned to the fermionic
spectrum. Each model build up on 331 symmetry has its own interesting properties according to the representations
of the gauge group used for the fermionic spectrum; that is the main reason to explore and identify the possible
sources of flavor changing neutral currents and lepton flavor violation at tree level. 
\end{abstract}

\newpage

\section{Introduction}

The Standard Model (SM) \cite{sm} has been successful to describe leptons, quarks and their interactions. But in any case, 
the SM leaves open questions concerning to the electroweak symmetry breaking sector of the model, as well as the 
particle content of the model: why there are three generations of quarks and leptons? These questions, among  others,
are the motivation to consider the SM as one important attempt to understand the elementary particles of nature and their
interactions but not to consider the SM as the ultimate theory of nature. A common alternative to look for new physics
beyond the SM is enlarging the gauge symmetry group, one of these alternatives is the gauge symmetry $SU(3)_C \times SU(3)_L \times
 U(1)_X$ (331) \cite{331}. There are many motivations for this new gauge symmetry group, one of them is that there are some of models based on
331 symmetry that explain why the family number must be three. This result is obtained from the anomaly free condition
which is satisfied when equal number of triplets and anti-triplets (taking into account the $SU(3)_C$) are present and requiring 
the sum of all fermion charges to vanish; but even that each generation is anomalous and the anomaly cancellation is given for 
three generations or multiply of three. Other motivation is concern to the feature that $\sin^2 \theta_W$ in this model should be less than $1/4$, it
is related to the ratio of the coupling constants $g'$ and $g$ of $U(1)_X$ and $SU(3)_L$,
$$
\left(\frac{g'}{g}\right)^2=\frac{\sin^2 \theta_W}{1-4 \sin^2 \theta_W}
$$
in this model there is an energy scale at which the perturbative character is lost and the energy scale is found using
the condition $\sin^2 \theta_W=1/4$ and it is order of $\sim 4 TeV$ \cite{Dias:2004dc}.

On the other hand, in the breaking symmetry of the 331 gauge symmetry to the gauge group of the SM and then to the $U(1)_{Q}$,
some new bosons appear such as a new neutral $Z'$ boson which is heavier than the SM gauge bosons and in all the 331 models
it can mediate flavor changing process at tree level. In contrast in the framework of the SM it is well known that flavor
changing neutral currents (FCNC) are strongly suppressed because they appear only at one loop level. Therefore, these
FCNC processes can help to put stringent bounds on the parameter space of these kind of models\cite{FCNC}. Our aim in this work
is to review the possible models that can be built based on the extended gauge symmetry 331 and identify the different
 sources of FCNC in the quark sector as well as the lepton sector. 

\section{331 models}

The gauge group to be consider is $SU(3)_C \otimes SU(3)_L \otimes U(1)_X$. Left handed particles are into $SU(3)_L$ 
triplets, there are the usual quarks and leptons plus new exotic particles, and the anomaly free condition constraints
the allowed fermion representations ($3$ or $3^*$) and the quantum numbers. To describe the particle content of 
the model and to identify specific types of 331 models is important how is defined the
electric charge operator, which can be written as a linear combination of
the diagonal generators of the group:
\begin{equation}
Q\ = \ T_3\,+\,\beta\, T_8\,+\,X\ ,
\label{qoperator}
\end{equation}
where $\beta$ is a parameter that characterizes the specific particle structure. The
parameter $\beta$ can be chosen $\beta=\pm \sqrt{3}$ or $\beta=\pm 1/\sqrt{3}$, obtaining 331 models with exotic 
electric charges or 331 models without exotic electric charges, by exotic charges we mean charges different from those
that appear in the SM framework. 

Since each lepton family has three states, taking massless neutrinos, they can be arranged into $SU(3)_L$ anti-triplets
$\psi_i^T =(l_i^- , -\nu_i , l_i^+)$ where $i$ is a family index. The first two components corresponds to 
the ordinary electroweak doublet. This model corresponds to $\beta=\sqrt{3}$ \cite{331} for the charge operator in equation
(\ref{qoperator}). Therefore each lepton family will be in the $(1,3^*)_0$ representation of $SU(3)_C \otimes SU(3)_L \otimes U(1)_X$.
With these assumptions there are no new leptons in the 331 model and all three lepton families are treated identically.
In contrast, one of the three quark families transform differently from the other two which is required to anomaly
cancellation. Anomaly cancellation requires that two families of quarks transform as triplets $(3,3)_{-1/3}$ and 
the third one transforms as an anti-triplet $(3,3^*)_{2/3}$. The right handed spectrum is put in singlets in the usual way 
$(3^*,1)_{-2/3,1/3,4/3}$ for the first two families and $(3^*,1)_{-5/3,-2/3,1/3}$ for the third one. It is worth 
to notice that in general the assumption that one quark family is transforming differently to the other two families is
 a general condition in the framework of 331 models and it is generally assumed that the unique generation corresponds
to the third generation and then it could explain the heavy top quark mass.

In the gauge sector, five new gauge bosons beyond the SM are found.  The new gauge bosons form a complex $SU(2)_L$ doublet
of dileptons $(Y^{++},Y^+)$ with hypercharge 3 and a singlet $W_\mu^8$. The gauge boson $W^8_\mu$ mixes with the gauge boson
$X$ from the $U(1)_X$ to form the hypercharge $B_\mu$ boson and a new neutral $Z'_\mu$ boson. 

In order to break the symmetry spontaneously, four Higgs multiplets are necessary. Three triplets in representations
$(1,3)_1$, $(1,3)_0$ and $(1,3)_{-1}$ for the breaking of $SU(2)\times U(1)$ in order to give masses to all quarks and
a sextet $(1,6)_0$ is required for the lepton masses \cite{331}. 

In this first model \cite{331} there are new sources of FCNC processes at tree level coming from the new $Z'$ boson
  in the 
quark sector because the families are treated differently. Also at one
loop level appears new contributions coming from the charged bileptons and the charged scalar sector \cite{ng}. In this model 
there are FCNC in the lepton sector and they are mediated by the charged bileptons\cite{ng}. 

A possible variation of this original model is to consider a new lepton assignment using  a heavy lepton $E^+$ instead
the $e^c$ and adding $e^c$ and $E^-$ as singlets \cite{tully}. With this model is easy to generate small neutrino masses and lepton number
violation can occur and one property of this model version is that bileptons only couple standard to exotic leptons \cite{tully}.

On the other hand is possible to obtain models based on the gauge 331 symmetry but without new exotic charges for the
fermions. One version of that is the model proposed by Ozer \cite{ozer} where is introduced a right handed neutrino. A systematic
study of these kind of models was done in references \cite{ponce}. According to the $\beta$ value in equation (\ref{qoperator}) is
possible to get six different set of fermions and the fermion structure in order to avoid the quiral anomalies producing
 different 331 models. The fermion sets are four lepton sets and two quark sets.

The first set of leptons is

\begin{center}
\begin{tabular}{|c|c|c|}
\hline
$L_1= \left(\begin{array}{c}
  \nu_i \\
  e_i^- \\
  E^-_i
\end{array}\right)$& $e^+_i$ & $E^+_i$ \\
  $(1,3)_{-2/3}$ & $(1,1)_1$ & $(1,1)_1$ \\ \hline
\end{tabular}
\end{center}
using $i=1,2,3$ as the family index and $e_i, d_i, u_i$ are the SM fermions and
$E_i, D_i, U_i$ are the exotic ones. 

The second set is 

\begin{center}
\begin{tabular}{|c|c|}
\hline
$L_2= \left(\begin{array}{c}
  e^-_i \\
  \nu_i \\
  N^0_i
\end{array}\right)$& $e^+_i$  \\
  $(1,3^*)_{-1/3}$ & $(1,1)_1$  \\ \hline
\end{tabular}
\end{center}
where there is a neutral exotic particle. For the third leptonic set,
\begin{center}
\begin{tabular}{|c|c|c|}
\hline
$L_3= \left(\begin{array}{c}
  e_i^- \\
  \nu_i \\
  N^0_1
\end{array}\right)$& $\left(\begin{array}{c}
  E_i^- \\
  N_2^0 \\
  N_3^0
\end{array}\right)$ & $\left(\begin{array}{c}
  N^0_4 \\
  E_i^+ \\
  e^+_i
\end{array}\right)$ \\
  $(1,3^*)_{-1/3}$ & $(1,3^*)_{-1/3}$ & $(1,3^*)_{2/3}$ \\ \hline
\end{tabular}
\end{center}
where there is a charged exotic particle and four new exotic neutral ones. Finally, for the forth set

\begin{center}
\begin{tabular}{|c|c|c|c|c|c|}
\hline
$L_4= \left(\begin{array}{c}
  \nu_i \\
  e^-_i \\
  E^-_{1i}
\end{array}\right)$& $\left(\begin{array}{c}
  E_{2i}^- \\
  N_1^0 \\
  N_2^0
\end{array}\right)$ & $\left(\begin{array}{c}
  N^0_3 \\
  E_{2i}^- \\
  E_{3i}^-
\end{array}\right)$ & $e^+_i$ & $E^+_{1i}$ & $E^+_{3i}$ \\
  $(1,3)_{-2/3}$ & $(1,3)_{1/3}$ & $(1,3)_{-2/3}$ & $(1,1)_1$ & $(1,1)_1$ & $(1,1)_1$ \\ \hline
\end{tabular}
\end{center}
with three exotic charged particles and three neutral.

Now, the quark sets are

\begin{center}
\begin{tabular}{|c|c|c|c|}
\hline
$Q_1= \left(\begin{array}{c}
  d_i \\
  u_i \\
  U_i
\end{array}\right)$& $d_i$ & $u_i$ & $U_i$ \\
  $(3,3^*)_{1/3}$ & $(3,1)_{1/3}$ & $(3,1)_{-2/3}$ & $(3,1)_{-2/3}$ \\ \hline
\end{tabular}
\end{center}
\begin{center}
\begin{tabular}{|c|c|c|c|}
\hline
$Q_2= \left(\begin{array}{c}
  u_i \\
  d_i \\
  D_i
\end{array}\right)$& $u_i$ & $d_i$ & $D_i$ \\
  $(3,3)_{0}$ & $(3,1)_{-2/3}$ & $(3,1)_{1/3}$ & $(3,1)_{1/3}$ \\ \hline
\end{tabular}
\end{center}
The anomaly contribution for each set is presented in table \ref{anomalias}

\begin{table}
\begin{center}
\begin{tabular}{||c||c|c|c|c|c|c||}
  \hline\hline
  Anomalies & $L_1$ & $L_2$ & $L_3$ & $L_4$ & $Q_1$ & $Q_2$
  \\\hline\hline
  $SU(3)_c^2U(1)_X$ & 0 & 0 & 0 & 0 & 0 & 0 \\\hline
  $SU(3)_L^2U(1)_X$ & -2/3 & -1/3 & 0 & -1 & 1 & 0 \\\hline
  $grav^2U(1)_X$ & 0 & 0 & 0 & 0 & 0 & 0 \\\hline
  $U(1)_X^3$ & 10/9 & 8/9 & 6/9 & 12/9 & -12/9 & -6/9 \\
  \hline\hline
\end{tabular}
\end{center}
\caption{Anomalies for the six fermion sets}
\label{anomalias}
\end{table}

Based on the table \ref{anomalias}, it is possible to build up many models asking
for the anomaly free condition. There are two one family models and eight three family models,
referring to how cancel out the anomalies if it is needed one family or the three families.
There are two one family models composed by the sets $Q_2+L_3$ and $Q_1+L_4$. These models were studied in references
\cite{ponce,ponce2} and their relation with the grand unified theories established. For the three family models, there are
the combinations $3L_2 + Q_1 + 2Q_2$,  $3L_1 + 2Q_1 + Q_2$, $2(Q2+L3)+(Q1+L4)$ and $2(Q1+L4)+(Q2 + L3)$ and there
are other two models particularly interesting because they treat the three family leptons completely different, they are
the combinations $L_1+L_2+L_3+Q_1+2Q_2$ and $L_1+L_2+L_4+2Q_1+Q_2$ \cite{ponce2,anderson}.

In the gauge sector there are  17 gauge bosons, one gauge boson $B^\mu$
associated to $U(1)_X$, eight gluons associated to $S(3)_c$ and eight gauge fields
from the $SU(3)_L$. The gauge bosons associated with $SU(3)_L$ transform according to the
adjoint representation of the group, and can be written as
\begin{equation}
\mathbf{W}_\mu \ = \ W_\mu ^a \frac{\lambda^a}{2} \ = \ \frac 12\left(
\begin{array}{ccc}
W_\mu^3 + \frac 1{\sqrt{3}}W_\mu ^8 & \sqrt{2}\, W_\mu ^{+} &
\sqrt{2}\, K_{1\mu} \\
\sqrt{2}\, W_\mu ^{-} & -W_\mu ^3 + \frac 1{\sqrt{3}} W_\mu ^8 &
\sqrt{2}\, \bar K_{2\mu} \\
\sqrt{2}\, \bar K_{1\mu} & \sqrt{2}\, K_{2\mu} & -\frac 2
{\sqrt{3}}W_\mu ^8
\end{array}
\right) \ ,
\label{3}
\end{equation}
where $\lambda^a$ are the Gell-Mann matrices, and the electric charges of
$K_1$ and $K_2$ are given by $Q_1 = 1/2+\sqrt{3}\beta/2$ and $Q_2 =
1/2-\sqrt{3}\beta/2$, respectively.

In general, it is convenient to rotate the neutral gauge bosons $W^3_\mu$,
$W^8_\mu$ and $B_\mu$ into new states $A_\mu$, $Z_\mu$ and $Z'_\mu$, given by
\begin{eqnarray}
\left(\begin{array}{c}
A_\mu \\ Z_\mu \\ Z_\mu^\prime
\end{array}\right) =
\left( \begin{array}{ccc}
S_W & \beta S_W & C_W\sqrt{1-\beta^2 T_W^2} \\
C_W & -\,\beta S_W T_W & -\,S_W\sqrt{1-\beta^2 T_W^2} \\
0 & -\sqrt{1-\beta^2 T_W^2} & \beta T_W
\end{array} \right)
\left( \begin{array}{c}
W_\mu^3 \\ W_\mu^8 \\ B_\mu
\end{array} \right) \ ,
\label{bosgauneu}
\end{eqnarray}
where the angle $\theta_W$ is defined by $T_W = \tan\theta_W
= g'/\sqrt{g^2+\beta^2{g'}^2}$, $g$, $g'$ being the coupling constants
associated to the groups $SU(3)_L$ and $U(1)_X$ respectively ($S_W =
\sin\theta_W$, etc.). In the new basis, $A_\mu$ (the photon) is the gauge
boson corresponding to the generator $Q$, while $Z_\mu$ can be identified
with the SM $Z$ boson. As in the SM, the extended electroweak symmetry is
spontaneously broken in 331 models by the presence of elementary scalars
having nonzero vacuum expectation values~\cite{Diaz:2003dk}. The symmetry breakdown follows a
hierarchy
\begin{equation}
SU(3)_L\otimes U(1)_X \stackrel{V} \rightarrow SU(2)_L\otimes U(1)_Y
\stackrel{v} \rightarrow\ U(1)_Q \ \ ,
\label{jerarq}
\end{equation}
in which two VEV scales $V$ and $v$, with $V\gg v$, are introduced. The
photon is kept as the only massless gauge boson, while the remaining neutral
gauge bosons get mixed. In this way, $Z$ and $Z'$ turn out to be only
approximate mass eigenstates.

\section{FCNC in 331 models}

First of all, the extension of the gauge group which embedded the SM group implies a new neutral gauge $Z'$ boson, which
in general in all the 331 models presented generates FCNC at tree-level. This fact is because in 331 models is not
possible to accommodate all the SM spectrum in multiplets with the same quantum numbers, therefore the $Z'$ couplings
are not universal for all the fermions and that is the origin of a new source of FCNC. Particularly,
to treat in a differ manner the third generation, as is usually assumed, to the other two generations produces
FCNC contributions. This property is common to all the 331 models in the quark sector. It is worth to mention that
even in the left handed couplings of the standard fermions to the $Z$ neutral boson appear FCNC at tree level through
the mixing of $Z-Z'$ and also coming from the mixing between the standard quarks and the exotic ones included 
in each case. Moreover, the mixing between neutral gauge bosons should take into account the gauge bosons which transform
according to the adjoint representation of the $SU(3)_L$, some noted $K^\pm$ and $K^0$ gauge bosons for 
the charged sector and neutral sector (they are related to $K_{1,2}$). 
In order to notice these effects clearly, the Lagrangian for the new $Z'$ boson with a $\beta$ 
arbitrary is the following
\begin{eqnarray}\label{corriente Zprima general}
{\cal L}^{Z^\prime}&=&-\frac{g^\prime}{2T_W} Z^{\mu\prime}\left[
\sum_{m=1}^2\overline{D^0}_{m}\gamma_\mu\left(\frac{P_L}{\sqrt{3}}
+\frac{T_W^2\beta}{3}(P_L-2P_R)\right)D_{m}^0\right.\nonumber \\
&+&\overline{D^0}_3\gamma_\mu
\left(-\frac{P_L}{\sqrt{3}}+\frac{T_W^2\beta}{3}(P_L-2P_R)\right)D_3^0\nonumber\\
&+&\sum_{m=1}^2\overline{U^0}_{m}\gamma_\mu\left(\frac{P_L}{\sqrt{3}}
+\frac{T_W^2\beta}{3}(P_L+4P_R)\right)U_{m}^0\nonumber\\
&+&
\overline{U^0}_3\gamma_\mu\left(-\frac{P_L}{\sqrt{3}}+\frac{T_W^2\beta}{3}(P_L+4P_R)\right)U_3^0\nonumber\\
&+&\overline{{L}^0}\gamma_\mu\left(-\frac{P_L}{\sqrt{3}}-T_W^2\beta(P_L+2P_R)\right){L}^0
+\overline{\nu^0}\gamma_\mu\left(-\frac{1}{\sqrt{3}}-T_W^2\beta
\right)P_L\nu^0\nonumber\\
&+&\sum_{m=1}^2\overline{J^0}_m \gamma_\mu\left(-\frac{2P_L}{\sqrt{3}}+T_W^2\left(\frac{1}{3}+
\frac{3\beta}{\sqrt{3}}\right)\right)J_m^0\nonumber\\
&+&\overline{J^0}_3\gamma_\mu\left(-\frac{2P_L}{\sqrt{3}}+T_W^2\left(\frac{1}{3}-
\frac{3\beta}{\sqrt{3}}\right)(P_L-P_R)\right)J_3^0\nonumber\\
&+&\left.+\overline{E^0}\gamma_\mu
\left(\frac{2P_L}{\sqrt{3}}+T_W^2\left(-\frac{1}{3}-
\frac{3\beta}{\sqrt{3}}\right)(-P_L+P_R)\right)E^{0}\right]\,.
\end{eqnarray}
where $D^0=\pmatrix{
d_1^0 & d_2^0 & d_3^0}^T$ , $U^0=\pmatrix{
u_1^0 & u_2^0 & u_3^0}^T$, ${L}^0=\pmatrix{
 e^0_1 & e^0_2 & e^0_3}^T$,  $E^{0}=\pmatrix{
 E_1^{0} & E_2^{0} & E_3^{0}}^T$ and the exotic quarks $j_i^0$ with electric charges given by  $q_{J_1}=Q_{J_2}=1/6+\sqrt{3}\beta/2$ and
$q_{J_3}=1/6-\sqrt{3}\beta/2$.
There is explicitly shown the no universal couplings between the quarks $D_i$, $U_i$ and the $Z'$ boson and it
is because one family is in the 3 representation while the other two are in the $3^*$ (or vice versa). As a consequence,
the FCNCs arise once the fields $U_i$ and $D_i$ are rotated to the mass eigenstates. The number of extra fermions
up-quark type or down-quark type depends on the parameter $\beta$, for $\beta=-1/\sqrt{3}$ will have $N_U=1$ and
$N_D=2$ and for  $\beta=1/\sqrt{3}$ will have $N_U=2$ and $N_D=1$. Therefore, there is not only FCNC at tree level through
the $Z'$ boson but also the usual $Z$ boson due to the mix of these new exotic quarks with the ordinary ones. To notice
this, for the case of $\beta=+1/\sqrt{3}$ the following definitions are useful 
$U_0^T=(u_1^0,u_2^0,u_3^0,T_1^0,T_2^0)$, $D_0^T=(d_1^0,d_2^0,d_3^0,B_1^0)$, 
$E_0^T=(e^0,\mu^0,\tau^0,E_1^0,E_2^0,E_3^0)$ and $N_0^T=(\nu_e^0,\nu_\mu^0,\nu_\tau^0)$. Meanwhile for the case 
$\beta=-1/\sqrt{3}$ the definitions are $U_0^T=(u_1^0,u_2^0,u_3^0,T_1^0)$, $D_0^T=(d_1^0,d_2^0,d_3^0,B_1^0,B_2^0)$, 
$E_0^T=(e^0,\mu^0,\tau^0)$ and $N_0^T=(\nu_e^0,\nu_\mu^0,\nu_\tau^0,,N_1^0,N_2^0,N_3^0)$. With this vector notation,
the Lagrangian for neutral currents is
\begin{eqnarray}\label{acopgenzz}
{\cal L}_{NC}&=&\sum_\Psi-\frac{gZ^\mu}{2C_W}\left\{\bar{\Psi^0}\gamma_\mu\epsilon^{(1)}_{\Psi_{(L)}}P_L{\Psi}^0
+\bar{\Psi^0}\gamma_\mu\epsilon^{(1)}_{\Psi_{(R)}}P_R{\Psi}^0 \right\}\nonumber\\
&&-\frac{g^\prime Z^{\prime\mu}}{2\sqrt{3}S_W C_W}\left\{\bar{\Psi^0}\gamma_\mu\epsilon^{(2)}_{\Psi_{(L)}}P_L{\Psi}^0
+\bar{\Psi^0}\gamma_\mu\epsilon^{(2)}_{\Psi_{(R)}}P_R{\Psi}^0 \right\}\nonumber\\
&&-\frac{g}{\sqrt{2}}\left\{\bar{\Psi^0}\gamma_\mu\epsilon^{(3)}_{\Psi_{(L)}}P_L{\Psi}^0 {\rm Re}K^\mu+i\bar{\Psi^0}\gamma_\mu\epsilon^{(4)}_{\Psi_{(L)}}P_L{\Psi}^0 {\rm Im}K^\mu\right\}
\end{eqnarray}
where the sum is over ${U_0}$, ${D_0}$, ${E_0}$ and ${N_0}$. The couplings $\epsilon^{(1,2)}_{\Psi_{(L,R)}}$
depends on the parameter $\beta$. With $\beta=\pm 1/\sqrt{3}$ the $Z^0$ interaction is
\begin{eqnarray}\label{acoupz}
\epsilon^{(1)}_{{\cal U}_{(L)_{}}}&=&\left(C_W^2-S_W^2/3\right){\bf 1}_{(3+N_U^\pm)\times (3+N_U^\pm)}-\pmatrix{
0_{(3\times3)} &  \cr  & {\bf 1}_{(N_U^\pm\times N_U^\pm)}}\,, \nonumber\\ \epsilon^{(1)}_{{\cal U}_{(R)_{}}}
&=&-\left(4S_W^2/3\right) {\bf 1}_{(3+N_U^\pm)\times (3+N_U^\pm)}\,.\nonumber\\
\epsilon^{(1)}_{{\cal D}_{(L)_{}}}&=&\left(-C_W^2-S_W^2/3\right){\bf 1}_{(3+N_D^\pm)\times (3+N_D^\pm)}+\pmatrix{
0_{(3\times3)} &  \cr  & {\bf 1} _{(N_D^\pm\times N_D^\pm)}}\,, \nonumber\\ \epsilon^{(1)}_{{\cal D}_{(R)_{}}}
&=&+\left(2S_W^2/3\right){\bf 1}_{(3+N_D^\pm)\times (3+N_D^\pm)}\,
\end{eqnarray}
where the no universality of the left handed quarks is clear while the right handed couplings drive for 
 $\epsilon^{(1)}_{{\cal U, D}_{(R)}}$ are universals.

In a similar way for the $Z^{\prime0}$ boson, the couplings are

\begin{eqnarray}\label{acopupzp}
\epsilon^{(2)}_{{\cal U}_{(L)_{}}}&=&\left(C_W^2\pm S_W^2/3\right){\bf 1}_{(3+N_U^\pm)\times (3+N_U^\pm)}
-2C_W^2\pmatrix{0_{(2\times2)} &   \cr  & {\bf 1}_{(N_T^\pm+1)\times (N_T^\pm+1)}}\nonumber\\
&+&(C_W^2\mp2C_W^2\pm S_W^2)\pmatrix{0_{(3\times3)} &   \cr  & {\bf 1}_{(N_U^\pm\times N_U^\pm)}}\,, \nonumber\\ 
\epsilon^{(2)}_{{\cal U}_{(R)_{}}}&=&\pm\frac{4S_W^2}{3} {\bf 1}_{(3+N_U^\pm)\times (3+N_U^\pm)}\,,\nonumber\\
\epsilon^{(2)}_{{\cal D}_{(L)_{}}}&=&\left(C_W^2\pm S_W^2/3\right){\bf 1}_{(3+N_D^\pm)\times (3+N_D^\pm)}
-2C_W^2\pmatrix{0_{(2\times2)} &   \cr  & {\bf 1}_{(N_D^\pm+1)\times (N_D^\pm+1)}}\nonumber\\
&+&(C_W^2\pm2C_W^2\mp S_W^2)\pmatrix{0_{(3\times3)} &   \cr  & {\bf 1}_{(N_D^\pm\times N_D^\pm)}}\,, \nonumber\\
\epsilon^{(2)}_{{\cal D}_{(R)_{}}}&=&\mp \left(2S_W^2/3\right) {\bf 1}_{(3+N_D^\pm)\times (3+N_D^\pm)}\, .
\end{eqnarray}

At this point is important to mention that the couplings in eqs. (\ref{acopgenzz})-(\ref{acopupzp}) are in the
interaction basis, thus to obtain the mass eigenstates is necessary to get the rotation matrices which diagonalize the
 mass matrices in the Yukawa sector. Therefore the mass eigenstates $U$ and $D$ are defined by
\begin{eqnarray} \label{automasaUpdown1}
U_L^0 = V_L^u \ U_L \,\,  , \,  \, D_L^0 = V_L^d D_L \ .
\end{eqnarray}
with matrices $V_L$ of dimensions $(3+N_U^\pm) \times (3+N_U^\pm)$ and $(3+N_D^\pm)\times (3+N_D^\pm)$ respectively.
It is useful to write the matrices $V_L^{u,d}$  as
\begin{equation}\label{matmezclaUpdown}
V_L^u \ = \
\left( \begin{array}{cc}
{V_0^u}_{(3\times 3)} & {V_X^u}_{(3\times N_U^\pm)} \\
{V_Y^u}_{(N_U^\pm\times 3)} & {V_U}_{(N_U^\pm\times N_U^\pm)}
\end{array} \right) \,\ {\rm and} \, \
V_L^d \ = \
\left( \begin{array}{cc}
{V_0^d}_{(3\times 3)} & {V_X^d}_{(3\times N_D^\pm)} \\
{V_Y^d}_{(N_D^\pm\times 3)} & {V_D}_{(N_D^\pm\times N_D^\pm)}
\end{array} \right) ,
\end{equation}
using submatrices in such a way that $V_{CKM} = V_0^{u\dagger}V_0^d$ and in general the CKM matrix is not unitary.

In addition, the models include new gauge bosons $K_\mu$ which coupled to the left handed fermions, the couplings 
in equation  (\ref{acopgenzz}) for the $K_2^\mu$ boson when $\beta=1/\sqrt{3}$ are
\begin{eqnarray}\label{acopKbpre}
\epsilon^{(3)}_{{ U}_{(L)_{}}} &=& \pmatrix{0_{2\times2} & & {\bf 1}_{2\times 2}\cr & 0 & \cr {\bf 1}_{2\times2} & & 0_{2\times 2}}
\qquad \epsilon^{(3)}_{{ D}_{(L)_{}}}=\pmatrix{0_{2\times2} &  & \cr  & 0 & 1 \cr & 1 & 0}\,,\nonumber\\
\epsilon^{(3)}_{{ E}_{(L)_{}}}&=&\pmatrix{& {\bf 1}_{3\times 3} \cr {\bf 1}_{3\times3} &}\,,\qquad
\epsilon^{(4)}_{{ U}_{(L)_{}}} = \pmatrix{0_{2\times2} & & {\bf 1}_{2\times 2}\cr & 0 & \cr -{\bf 1}_{2\times2} & & 0_{2\times 2}}
\,\nonumber\\
\epsilon^{(4)}_{{ D}_{(L)_{}}}&=&\pmatrix{ 0_{2\times2} &  & \cr  & 0 & -1 \cr & 1 & 0}\,
\qquad\epsilon^{(4)}_{{ E}_{(L)_{}}}=\pmatrix{
& {\bf 1}_{3\times 3} \cr -{\bf 1}_{3\times3} &}\,.
\end{eqnarray}

And when $\beta=-1/\sqrt{3}$ for the $K_1^\mu$ boson, they are
\begin{eqnarray}\label{acopKbmre}
\epsilon^{(3)}_{{ U}_{(L)_{}}}&=&\pmatrix{0_{2\times2} &  & \cr  & 0 & 1 \cr & 1 & 0}
\qquad \epsilon^{(3)}_{{ D}_{(L)_{}}}=\pmatrix{
0_{2\times2} & & -{\bf 1}_{2\times 2}\cr & 0 & \cr -{\bf 1}_{2\times2} & & 0_{2\times 2}}\nonumber\\
\epsilon^{(3)}_{{ N}_{(L)_{}}}&=&\pmatrix{& {\bf 1}_{3\times 3} \cr {\bf 1}_{3\times3} &}\qquad
\epsilon^{(4)}_{{ D}_{(L)_{}}}=\pmatrix{0_{2\times2} &  & \cr  & 0 & 1 \cr & -1 & 0}\nonumber\\
\epsilon^{(4)}_{{ U}_{(L)_{}}}&=&\pmatrix{0_{2\times2} & & {\bf 1}_{2\times 2}\cr & 0 & \cr -{\bf 1}_{2\times2} & & 0_{2\times 2}}
\qquad\epsilon^{(4)}_{{ N}_{(L)_{}}}=\pmatrix{& {\bf 1}_{3\times 3} \cr -{\bf 1}_{3\times3} &}\,.
\end{eqnarray}

It is often assumed that the vacuum expectation values of the scalar fields are real, this assumption implies
that there is not any spontaneous CP symmetry breaking. In that case, the state $Im K$ decouples and therefore
turn into an exact mass eigenstate. However, the bosons $Z^\mu$, $Z^{\prime\mu}$ and $\sqrt{2}{\rm Re}K$ mix and it 
is possible to get the mass basis $(Z_1,Z_2,Z_3)$ through an orthogonal matrix which depends on the vacuum expectation
values of the Higgs bosons,
\begin{eqnarray}
\pmatrix{
Z \cr Z^\prime \cr \sqrt{2}{\rm Re}K}= R \pmatrix{
Z_1 \cr Z_2 \cr Z_3}
\end{eqnarray}
and therefore the Lagrangian can be re-written as
\begin{eqnarray}
{\cal L}_{NC} & = & - \sum_{\Psi = {\cal U},{\cal D}}
\left[ Q_\Psi \bar{\Psi} \gamma^\mu \Psi \, A_\mu \; + \sum_{j,k=1}^3
g_j \bar{\Psi} \gamma^\mu (E_{\Psi_L}^{(j)} P_L +
E_{\Psi_R}^{(j)}P_R) \Psi \, R_{jk}\, Z_{k\mu}\right. \nonumber \\
& & \left.\ \ \ \ \ \ \ \ \ \ \ \ + \; i\,\frac{g}{2}
\bar{\Psi} \gamma^\mu (E_{\Psi_L}^{(4)} P_L +
E_{\Psi_R}^{(4)}P_R) \Psi \, \sqrt{2}\,{\rm Im}K_\mu \right] \ ,
\label{neucurr}
\end{eqnarray}
where $Q_\Psi$ is the electric charge and the coupling constants $g_j$ are 
\begin{equation}
g_1 = \frac{g}{2C_W} \ ,\qquad
g_2 = \frac{g'}{2\sqrt{3}S_WC_W} =
\frac{g}{2\sqrt{3}C_W\sqrt{C_W^2 - \beta^2 S_W^2}}\ ,\qquad
g_3 = \frac{g}{2} \ ,
\end{equation}
and the matrices $E^{(i)}_{\Psi_{L,R}}$ are given by
\begin{equation}
E_{\Psi_L}^{(i)} \ = \ V_L^{\Psi\dagger} \epsilon^{(i)}_{\Psi_L} V_L^\Psi
\ , \qquad \qquad
E_{\Psi_R}^{(i)} \ = \ V_R^{\Psi\dagger} \epsilon^{(i)}_{\Psi_R} V_R^\Psi
\ = \ \epsilon^{(i)}_{\Psi_R} \ .
\end{equation} 

Finally about the sources of FCNC in the framework of the 331 models, they are two models which are very interesting
because they have some special properties from the phenomenological point of view. They are the models build up
with the fermionic sets $L_1+L_2+L_3+Q_1+2Q_2$ and $L_1+L_2+L_4+2Q_1+Q_2$. They not only differentiate the quark generations, doing one family
specially different, but they also do in the leptonic sector. These models will have the usual FCNC at tree level in 331
models in the quark sector through the $Z'$ boson but also they present FCNC in the leptonic sector through the
 scalar fields
and through the $Z'$ boson \cite{anderson}. To notice the new sources of FCNC arising in these models, the neutral current Lagrangian 
is going to be obtained. First of all, the spectrum should be specified
\begin{eqnarray}
\ell_{1L}=\pmatrix{\nu_1 \cr e_1^-\cr E_1^-}_L\,,\qquad \ell_{mL}=\pmatrix{e_m^-\cr \nu_m \cr N_k^0}_L\,,\qquad \ell_{5L}=\pmatrix{E_2^-\cr N_3^0\cr N_4^0}_L \,, \qquad \ell_{4L}=\pmatrix{N_5^0\cr E_2^+\cr e_3^+}_L
\end{eqnarray}
where $m=2,3$, $k=1,2$ and note that one of the leptonic triplets is in the adjoint representation respect to the other 
two then FCNC at tree level will arise through the $Z'$ boson. Using vector notation, the neutral current Lagrangian
in this case is
\begin{eqnarray}\label{lagrangiano}
{\cal L}_{NC}&=&\sum_\Psi\left[A_\mu\left\{\bar{\Psi^0}\gamma_\mu\epsilon^{A}_{\Psi_{(L)}}P_L{\Psi}^0
+\bar{\Psi^0}\gamma_\mu\epsilon^{A}_{\Psi_{(R)}}P_R{\Psi}^0 \right\}\right.\\\nonumber
&+&\frac{gZ^\mu}{2C_W}\left\{\bar{\Psi^0}\gamma_\mu\epsilon^{Z}_{\Psi_{(L)}}P_L{\Psi}^0
+\bar{\Psi^0}\gamma_\mu\epsilon^{Z}_{\Psi_{(R)}}P_R{\Psi}^0 \right\}\nonumber\\
&+&\left.\frac{g^\prime Z^{\prime\mu}}{2\sqrt{3}S_W C_W}\left\{\bar{\Psi^0}\gamma_\mu\epsilon^{Z'}_{\Psi_{(L)}}P_L{\Psi}^0
+\bar{\Psi^0}\gamma_\mu\epsilon^{Z'}_{\Psi_{(R)}}P_R{\Psi}^0 \right\}\right]
\end{eqnarray}
\noindent Defining the vector ${ E^T}=\left(e_1^-,  e_2^-,  e_3^-, E_1^-,  E_2^-\right)$, the couplings are

\begin{eqnarray}
\epsilon_{ { E}_{(L)}}^{A} &=& g S_W I_{5\times5}\,\,,\,\,\epsilon_{{\cal E}_{(R)}}^{A}=g S_W I_{5\times5}  \\
\epsilon_{{ E}_{(L)}}^{\it Z} &=& \frac{g}{2C_W}\,\mathtt{Diag}(C_{2W},C_{2W},C_{2W},-2S_W^2,C_{2W})\nonumber\\
\epsilon_{{ E}_{(R)}}^{\it Z} &=& \frac{g}{2C_W}\,\mathtt{Diag}(-2S_W^2,-2S_W^2,-2S_W^2,-2S_W^2,C_{2W})\nonumber\\
\epsilon_{{ E}_{(L)}}^{\it Z'} &=& \frac{g'}{2\sqrt{3}S_W C_W} \mathtt{Diag}(1,-C_{2W},-C_{2W},-C_{2W},-C_{2W}) \nonumber\\
\epsilon_{{ E}_{(R)}}^{\it Z'} &=& \frac{g'}{2\sqrt{3}S_W C_W} \mathtt{Diag}(2S_W^2,2S_W^2,-C_{2W},2S_W^2,1)\nonumber
\end{eqnarray}
Where  $C_{2W}=\cos{(2\theta_W)}$ and it is worthwhile to point out that the right handed couplings are not universal and it is a new feature
of this model. Usually in the framework of the 331 models only the left handed couplings are not universal 
but the right handed are universal as it was shown in equation (\ref{corriente Zprima general}). 

For the neutral sector  ${ N^T}=\left(\nu_1^0,  \nu_2^0,  \nu_3^0,  N_1^0,  N_2^0,  N_3^0,  N_4^0\right)$ is defined
and the left handed couplings are 
\begin{eqnarray}
\epsilon_{{ N}_{(L)}}^{A}&=&0 \, \,,\,\,
\epsilon_{{ N}_{(L)}}^{Z} = \frac{g}{2C_W}\,\mathtt{Diag}(1,1,1,0,0,1,0,-1)  \\
\epsilon_{{ N}_{(L)}}^{Z} &=& \frac{g'}{2\sqrt{3}S_WC_W} \mathtt{Diag}(1,-C_{2W},-C_{2W},2C_W^2,2C_W^2,-C_{2W},2C_W^2,-1)\nonumber
\end{eqnarray}

\section{Summary}

One of the most intriguing options to consider physics beyond the SM consists in to extend the gauge symmetry group
to  $SU(3)_C \times SU(3)_L \times U(1)_X$. There are many models based on the 331 symmetry and one intriguing
feature of these models are the presence of FCNC at tree level but the source of that new interactions is not unique
and depends on how the model is built up. In the Pleitez-Frampton model, the first one proposed, was established
the presence of FCNC at tree level coming from the new $Z'$ boson and due to the different assignment of the quark
representation for one of the quark families; doing the left handed couplings between quarks and the $Z'$ 
boson not universal. On the other hand, it is possible to build up models based on 331 symmetry contrary to the 
Pleitez-Frampton model without any exotic charges for the new particles in the spectrum. These kind of models 
correspond to a $\beta=\pm 1/\sqrt{3}$ in the electric charge operator (\ref{qoperator}). These models include
new exotic up-quark type and down-quark type which are going to mix with the standard quarks. In one version appears
five up quark type and four down quark type and another version include four up quark type and five down quark type;
also these models include extra charged leptons in one case and neutral leptons in the other one. The mixing obtained
is a source of FCNC at tree level when the quark fields are written in the mass basis. There are also a new source 
of FCNC which is coming from the mixing in the gauge sector between the bosons $(A,Z,Z',K)$. The mixing in 
this sector is usually reduced to the mixing between $Z$ and $Z'$. If it is consider the mixing between the quarks and
the mixing $(Z, Z')$ then the FCNC interactions appear through the $Z$ and the $Z'$ mediation. In the leptonic sector
something similar is going to happen. Finally there are models which not only treat different the quark families but the
leptonic families too. One of these models is presented and the neutral current Lagrangian obtained and one interesting
new and additional feature is the non universal couplings in the right handed sector through the $Z$ and $Z'$ bosons.
This new contributions to the FCNC processes could help to relax the bounds obtained on the $Z'$ Boson mass.\\\\ 
We thank to Prof. V. Pleitez for useful discussions. J.D.  acknowledges 
IFT-Universidade Estadual Paulista the financial support and J.A.R for its hospitality 
while this work was finished.

\end{document}